\documentclass[11pt,a4paper]{article}
\pdfoutput=1
\usepackage[utf8]{inputenc}

\usepackage{amsmath,amssymb}
\usepackage{epsfig,graphicx}
\usepackage{subfigure}
\usepackage{graphicx}
\usepackage{rotating}
\usepackage{cancel}
\usepackage{bm}
\usepackage{color}
\usepackage{comment}
\usepackage{cite}
\usepackage{psfrag}
\usepackage[pagebackref]{hyperref}

\newcommand{\diff}{\mathrm{d}}
\newcommand{\half}{\frac{1}{2}}
\newcommand{\me}{\mathrm{e}}

\renewcommand\({\left(}
\renewcommand\){\right)}
\renewcommand\[{\left[}

\newcommand{\exclude}[1]{}

\def\beq{\begin{equation}}
\def\eeq{\end{equation}}

\usepackage{textpos}

\begin{document}
\numberwithin{equation}{section}
\title{
\vspace{2.5cm} 
\Large{\textbf{On the Wondrous Stability of ALP Dark Matter
\vspace{0.5cm}}}}

\author{Gonzalo Alonso-\'Alvarez$^{1}$, Rick S. Gupta$^{2}$, Joerg Jaeckel$^{1}$ and Michael Spannowsky$^{2}$\\[2ex]
\small{\em $^1$Institut f\"ur theoretische Physik, Universit\"at Heidelberg,} \\
\small{\em Philosophenweg 16, 69120 Heidelberg, Germany}\\[0.5ex]
\small{\em $^2$Institute for Particle Physics Phenomenology,} \\
\small{\em South Road, Durham DH1 3LE, United Kingdom}\\[0.8ex]
}

\date{}
\maketitle
\begin{textblock*}{3cm}(12.3cm,-10cm)
	IPPP/19/84
\end{textblock*}

\begin{abstract}
\noindent
The very low mass and small coupling of axion-like particles (ALPs) is usually taken as a guarantor of their cosmological longevity, making them excellent dark matter candidates. That said, Bose enhancement could stimulate decays and challenge this paradigm. Here, we analyze and review the cosmological decay of ALPs into photons, taking Bose enhancement into account, thereby going beyond the usual naive perturbative estimate.
At first glance, this calculation seems to yield an exponentially growing resonance and therefore an extremely fast decay rate. However, the redshifting of the decay products due to the expansion of the Universe as well as the effective plasma mass of the photon can prevent an efficient resonance. While this result agrees with existing analyses of the QCD axion, for more general ALPs that can feature an enhanced photon coupling, stability is only ensured by a combination of the expansion and the plasma effects.
\end{abstract}

\newpage

\section{Introduction}
Axions, axion-like particles (ALPs), as well as other very light, weakly coupled bosons are promising and popular dark matter candidates~\cite{Abbott:1982af,Preskill:1982cy,Dine:1982ah,Arias:2012az}. Usually lacking a conserved charge\footnote{See~\cite{Alonso-Alvarez:2019pfe} for an example of very light dark matter particles featuring a conserved charge.}, they can in principle decay, and generally do. However, both their weak couplings as well as their low mass contribute to the decay rate being extremely small, and decays are often of little concern to the model builder.

\begin{figure}[!t]
\centering
\includegraphics[width=0.8\textwidth]{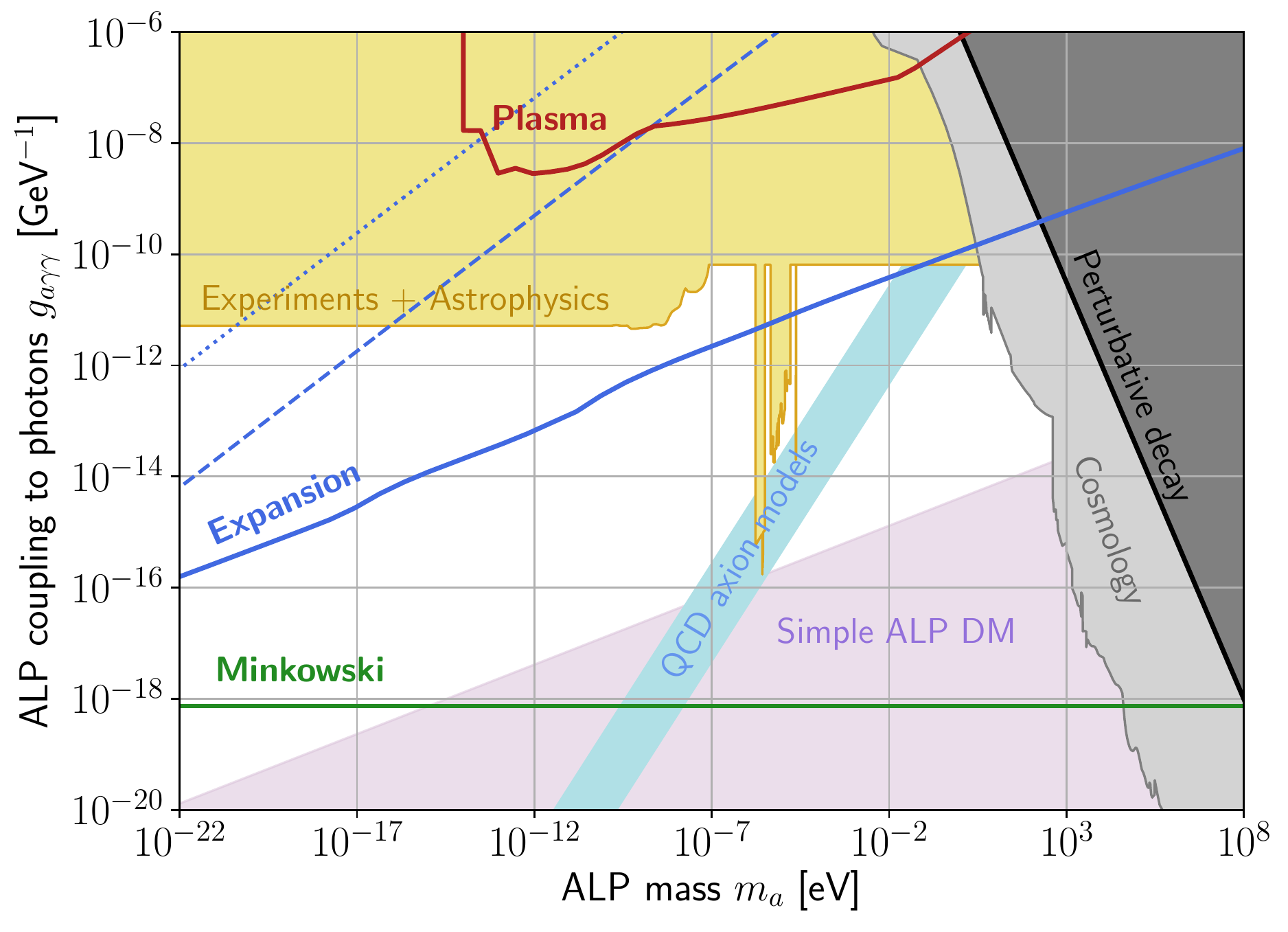}
\hspace*{1cm}
\caption{Parameter space for ALPs in the mass vs. two-photon coupling plane (adapted from~\cite{Redondo:2008en,Jaeckel:2010ni,Arias:2012az,Irastorza:2018dyq}). The shaded regions indicate experimental and observational limits as labelled.
The green line indicates the stability bound that would arise from Bose enhanced decay, neglecting expansion and plasma effects. The blue lines shows how the expansion of the Universe ameliorates this constraint. The dotted line corresponds to evaluating the stability condition today, the dashed one corresponds to matter-radiation equality and the solid line is obtained by evaluating it at the earliest possible time, when the field just starts to oscillate, i.e. when $H\sim m_{\phi}/3$. Finally, the red line additionally takes into account the plasma effects that prevent decay. We conclude that stability is ensured in all regions that are not already excluded by experiments or observations.
}
\label{fig:master}	
\end{figure}

Let us illustrate the situation at hand with the paradigmatic example of an ALP $\phi$ with a two-photon coupling,
\begin{equation}
{\mathcal{L}}=\frac{1}{2}\left(\partial_{\mu}\phi \right)^2 -\frac{1}{2}m^2_{\phi}\phi^2-\frac{1}{4}g_{\phi\gamma\gamma}\phi  F^{\mu\nu}\tilde{F}_{\mu\nu}.
\end{equation}
The perturbative decay rate is then given by 
\begin{equation}
\label{eq:pert}
\Gamma_{\rm pert}=\frac{g^{2}_{\phi\gamma\gamma}m^{3}_\phi}{64\pi}.
\end{equation}
The region where $\phi$ becomes cosmologically unstable corresponds to $\Gamma_{\rm pert}>H_0$, where $H_0$ is the present-day Hubble parameter, and is indicated as the solid black line labelled ``Perturbative decay" in Fig.~\ref{fig:master}.
As expected, the decay is only relevant for large masses $\gtrsim 1\,{\rm keV}$.

The estimate above showcases the fact that the interactions are very small and the corresponding rate is tiny. So why have another look and reconsider the cosmological relevance of decay processes?
The reason lies in the generation of the dark matter ALPs, which, as is well known, are produced coherently via the misalignment mechanism~\cite{Abbott:1982af,Preskill:1982cy,Dine:1982ah}. The key feature to highlight is the coherent nature of the produced particles.
Indeed, if the ALP is present during inflation, the field is completely homogenized within the horizon. Therefore, all ALPs are in the same state, i.e. their occupation number is enormous,
\begin{equation}
N_\phi\sim \frac{\rho_\phi}{m_\phi H^{3}}\sim 10^{40} \left(\frac{\rm eV}{m_{\phi}}\right)^2\left(\frac{\Phi}{10^{11}\,{\rm GeV}}\right)^2\left(\frac{m_{\phi}}{H}\right)^3,
\end{equation}
where $H=\dot{a}/a$ denotes the Hubble constant, $m_{\phi}$ the mass of the ALP field and $\Phi$ the typical field amplitude.
Such a large occupation number makes Bose enhancement effects seem plausible.
Yet, this by itself does not necessarily result in an enhanced decay rate. Bose enhancement also requires large occupation numbers in the final state, in our case populated by photons. Interestingly, since the decay into two photons is a two body decay and therefore all produced photons have equal energy, they can quickly accumulate a high occupation number, allowing for Bose enhancement to set in. 
As we will see in more detail below, it is here that the crux of the matter, which saves the perturbative estimate in all standard cases, lies. In the expanding Universe, the produced photons redshift, thereby moving out of the momentum range required for them to take part in the Bose enhancement and generally making ALPs cosmologically stable. Furthermore, the fact that the  Universe is permeated by an ionized plasma that generates an effective mass for the photons also contributes to hinder the decay. 

The problem of the potential instability of coherent bosonic dark matter is as old as axion dark matter itself~\cite{Abbott:1982af,Preskill:1982cy,Dine:1982ah}. Indeed, the authors of~\cite{Abbott:1982af,Preskill:1982cy} already realized that coherent decay of QCD axion dark matter into either relativistic axions or photons is ineffective due to expansion and plasma effects -- the two effects that we will reconsider here for more general ALPs. A non-equilibrium QFT calculation of the photon production from coherent QCD axion dark matter oscillations was performed in~\cite{Lee:1999ae} and reached the same conclusion, albeit plasma effects were not included in the computation. The decay of the condensate into relativistic axion modes can also be described as a growth of quantum fluctuations caused by anharmonicities in the potential. This was done in~\cite{Abbott:1982af,Kolb:1998iw}, once more finding that the effect is negligible in an expanding background, unless extreme misalignment angles occur~\cite{Arvanitaki:2019rax}.

As the situation is well-stablished in the case of QCD axions, in this work we consider the more general case of axion-like particles, where the mass and the decay constant (and therefore the photon coupling) are independent parameters.
The question of stability is of particular interest for ALPs with larger couplings and/or larger initial field values, which are theoretically motivated by
monodromies~\cite{McAllister:2008hb,Jaeckel:2016qjp}, clockwork constructions~\cite{Choi:2014rja,Choi:2015fiu,Kaplan:2015fuy} and string theory expectations~\cite{Halverson:2019cmy}. On the more experimental side, the $(m_\phi,\,g_{a\gamma\gamma})$ parameter space will be tested by many future ALP dark matter experiments such as ABRACADABRA~\cite{Kahn:2016aff}, ADMX~\cite{Shokair:2014rna},  BRASS~\cite{brasscite}, CULTASK~\cite{Petrakou:2017epq} HAYSTAC~\cite{Droster:2019fur}, KLASH~\cite{Gatti:2018ojx}, MadMax~\cite{Majorovits:2017ppy}, ORGAN~\cite{McAllister:2017lkb}, QUAX~\cite{Alesini:2019ajt}, or RADES~\cite{Melcon:2018dba}. In addition, experiments like IAXO~\cite{Irastorza:2013dav} will look for ALPs with photon couplings independently of whether they constitute the dark matter.
As we will see, it is a non-trivial conspiracy of expansion and plasma effects that allow the ALP to be sufficiently stable and therefore a viable dark matter candidate in the region of parameter space probed by these experiments.

Before returning to the main topic of this note, let us point out that even if the cosmological stability is not threatened, interesting effects related to ALP-photon interactions can take place. For instance, decays in dark matter-rich environments can lead to observable radio signals~\cite{Caputo:2018ljp,Caputo:2018vmy}; while photon propagation in an axion background can exhibit parametric resonance-driven conversions~\cite{Espriu:2011vj, Yoshida:2017ehj}. The associated signals would be enhanced if axion structures exist in our galaxy, be it in the form of axion stars~\cite{Tkachev:1986tr, Tkachev:1987cd, Tkachev:1987ci}, miniclusters~\cite{Kephart:1986vc, Kephart:1994uy, Tkachev:2014dpa, Hertzberg:2018zte, Sawyer:2018ehf}, condensates~\cite{Arza:2018dcy, Sigl:2019pmj}, or superradiant clouds around black holes~\cite{Rosa:2017ury, Sen:2018cjt, Ikeda:2019fvj}. On a different direction, it has been recently suggested that dark matter axions could be detected by looking for an electromagnetic ``echo"~\cite{Arza:2019nta}. Finally, the ALP condensate could also suffer similar tachyonic or parametric instabilities that can lead to a rapid decay into dark photons. The authors of~\cite{Agrawal:2017eqm} argued that this effect could help deplete an excess of dark matter axions. This idea was further developped in~\cite{Kitajima:2017peg}, while~\cite{Agrawal:2018vin, Co:2018lka} used the instability to generate dark photon dark matter\footnote{See~\cite{Arias:2012az,Graham:2015rva, Dror:2018pdh, Bastero-Gil:2018uel, AlonsoAlvarez:2019cgw, Ema:2019yrd, Nakayama:2019rhg} for other scenarios for the production of light vector dark matter.}. This violent decay process can lead to the production of gravitational waves, as was pointed out in~\cite{Machado:2018nqk}.

In this work we focus on the effect of a photon coupling, as it is the only bosonic Standard Model particle that is lighter than the typical dark matter ALPs. However, modifications of the simplest pseudo-Goldstone scenario for ALPs can also increase the effect of other couplings. An important example are the self couplings. While these effects are small for simple pseudo-Goldstones\cite{Abbott:1982af,Kolb:1998iw}, they can become important in a monodromy setup where the initial field values are large~\cite{Jaeckel:2016qjp}. This can lead to a fragmentation of the condensate and a significant density fluctuations at small scales~\cite{Berges:2019dgr}. Moreover, if the ALP exhibits couplings to other light bosonic dark sector particles, similar effects to what we discuss in the following may become relevant.

This brief note presents a detailed study of the cosmological stability of generic light pseudoscalar dark matter. The discussion in the previous paragraphs sets the plan for the rest of the manuscript. In the following section~\ref{sec:bose}, we use classical equations of motion to account for coherent effects, i.e. Bose enhancement, in the decay. This naively suggests an enormously enhanced decay rate. Then, in sections~\ref{sec:expansion} and~\ref{sec:plasma_mass}, the effects of expansion and in-medium photon dispersion are consecutively taken into account. This generally prevents the Bose enhanced regime from being reached.
Finally, we conclude in section~\ref{sec:conclusions}. 

\section{Bose enhanced decay rates}\label{sec:bose}
An alternative to the usual perturbative calculation of the decay rate $\phi\rightarrow \gamma\gamma$ is to consider the growth of the photon field in the time-dependent ALP background. We first study this in Minkowski space where the alternative approach may be expected to yield the same result as the perturbative one. The equation of motion (EoM) in Fourier space reads (see, e.g.~\cite{Agrawal:2017eqm})
\begin{equation}
\label{eq:motion}
\ddot{A}_{\pm}+\left(k^2\pm g_{\phi\gamma\gamma}\,k\,\dot{\phi}\right)A_{\pm}=0,
\end{equation}
where $k$ is the momentum of the photon mode and $A_{\pm}$ represent the two helicity modes for the photon.
For brevity, we will consider only the mode $A\equiv A_{-}$, but $A_{+}$ can be treated analogously.
The ALP dark matter field, which is taken to be homogeneous, performs oscillations with frequency $m_\phi$ and amplitude $\Phi$ and is therefore given by
\begin{equation}
\phi(t)=\Phi\sin(m_\phi t),
\end{equation}
corresponding to the solution to its free EoM\footnote{In our calculation, we neglect the backreaction of the decay photons on the ALP field. This is a good approximation as long as only a small number of photons is produced. Once photon production becomes significant enough for backreaction effects to matter, the stability is already compromised.}. Using the definitions
\begin{equation}
x=\frac{m_\phi}{2}t,\qquad \lambda=\frac{4k^2}{m^2_\phi},\qquad{\rm and}\qquad q=2g_{\phi\gamma\gamma}\Phi\frac{k}{m_\phi},
\end{equation}
the EoM of the photon field has the form of a standard Mathieu equation~\cite{Mathieu},
\begin{equation}
\frac{\diff A}{\diff x}+(\lambda-2q\cos(2x))A=0.
\label{mathieu}
\end{equation}
The crucial result is that this differential equation admits exponentially growing solutions
\begin{equation}
A\sim \exp\left(\eta_{k} t\right),
\end{equation}
with a rate \begin{equation}
\label{eq:growth}
\eta_{k}= {\beta_k\, g_{\phi\gamma\gamma} \Phi\, m_\phi},
\end{equation}
where $\beta_{k}$ is an ${\mathcal{O}}(1)$ coefficient.
Such growing solutions appear for $k$-modes satisfying the condition
\begin{equation}
\label{eq:phasespace}
\left| k-\frac{m_\phi}{2} \right| < \delta k,
\end{equation}
where the effective width of the resonance band is (see e.g.~\cite{kovacic2018} for an analytical derivation)
\begin{equation}
\label{eq:width}
\delta k \sim g_{\phi\gamma\gamma} \Phi\, m_\phi/2.
\end{equation}
The coefficient $\beta_k$ reaches its maximum $\beta_{k_\star}=1/4$ at the center of the band $k=k_\star=m_\phi/2$, and decreases to zero at the edges. This is nothing but the well-known phenomenon of parametric resonance~\cite{Traschen:1990sw,Kofman:1994rk,Shtanov:1994ce,Boyanovsky:1996sq,Kofman:1997yn,Berges:2002cz}.

The (dominant\footnote{The Mathieu equation also features a number of sub-dominant resonances. These are, however, parametrically weaker~\cite{kovacic2018}.}) resonance occurs around $k_\star=m_\phi/2$ as one would expect for a perturbative two-photon decay. The puzzling observation is however that both the width, $\delta k$, and the effective growth rate of the photon field, $\eta_{k}$, given in Eq.~\eqref{eq:growth}, are parametrically different from the perturbative decay rate given in Eq.~\eqref{eq:pert}. 
Crucially, the rate obtained from the EoM is proportional to only one power of the coupling constant, and it also depends on the ALP field amplitude and therefore on the ALP density. 
For example, let us consider the QCD axion with $f=10^{12}$ GeV, for which the perturbative decay time
\begin{equation}
\Gamma^{-1}_{\rm pert}\sim10^{51}\ \mathrm{s}
\end{equation}
is comfortably larger than the age of the Universe. However, the decay time obtained from the EoM, 
\begin{equation}
\eta_{k_\star}^{-1}\sim 10^{-7}\ \mathrm{s},
\end{equation} 
is dramatically smaller. For a generic ALP, the region of parameter space where the lifetime obtained from the EoM in Eq.~\eqref{eq:motion} is smaller than a Hubble time is indicated by the green line labelled ``Minkowski" in Fig.~\ref{fig:master}. The enhanced rate thus seems to affect huge areas in parameter space. However, as we will see in the next section, this instability does not become effective in a large part of this region once we take proper account of the expansion of the Universe and the effective plasma mass of the photon.

Before going on to discuss these effects, let us nevertheless try to understand the origin of the huge discrepancy between the classical and the standard perturbative calculations.
As already mentioned in the introduction, we are dealing with bosonic systems featuring very large occupation numbers. In such a situation (cf., e.g.,~\cite{Baumann}) the decay rate receives a Bose enhancement via the photon occupation number $N_k$,
\begin{equation}
\Gamma=\Gamma_{\rm pert}(1+2N_k).
\end{equation}
For $N_k\gg 1$, the second term can lead to an exponential growth in the photon number. This can easily be seen by writing the Boltzmann equation for photons with momentum $k_\star = m_\phi/2$ (we denote number densities by lowercase $n$ and occupation numbers by uppercase $N$),
\beq
\dot{n}_\gamma=2 \Gamma n_\phi=2 \Gamma_{\rm pert} (1+ 2 N_k) n_\phi.
\label{eq:boseen}
\eeq
This rate equation can be rewritten using the relations
\begin{equation}
\Gamma_{\rm pert}=\frac{g_{\phi\gamma\gamma} ^2 m_\phi^3}{64 \pi} \, , \quad n_\phi=\half m_\phi \Phi^2 \, , \quad \mathrm{and} \quad N_k \sim \frac{n_\gamma / 2}{4 \pi k_\star^2 \mathop{\delta k}/(2 \pi)^3} \, ,
\label{eq:expressions}
\end{equation}
where we have divided the photon number density by the corresponding phase-space volume to obtain the expression for the photon occupation number. 
For now we will simply insert Eq.~\eqref{eq:width} for the width of the momentum shell $\mathop{\delta k}$ (which is derived from the EoM Eq.~\eqref{eq:motion}). Below we will justify this from a perturbative argument based on the uncertainty principle.
Doing this, we can solve Eq.~\eqref{eq:boseen} in the $N_k\gg1$ limit to obtain a number density
\begin{equation}
\label{eq:photonevol}
n_\gamma \propto \exp{\left(\beta  g_{\phi\gamma\gamma}\Phi m_\phi t\right)},
\end{equation}
and so the result from the classical EoM is recovered.
The fudge factor $\beta$ takes into account the ${\cal O}(1)$ uncertainty in the expression for $N_k$ in Eq.~\eqref{eq:expressions}. 

The assumption that $N_k\gg1$ needs justification, as initially no photons may be present in the relevant phase-space region. Let us estimate the time needed for the photon occupation number to grow beyond $1$ via the perturbative decay. To evaluate this, we start from Eq.~\eqref{eq:boseen} in the regime where $N_k\ll1$, and therefore ignore the second term to obtain
\beq
\Delta n_\gamma= 2 \Gamma_{\rm pert} n_\phi \Delta t.
\label{eq:neq1}
\eeq
If the decay happens within a short time $\mathop{\Delta t}$, the momentum of the photons has an indetermination $\mathop{\Delta k}$ due to the uncertainty principle $\mathop{\Delta E}\mathop{\Delta t}\geq 1/2$. Therefore, the corresponding growth in the occupation number after a time $\mathop{\Delta t}$ is
\begin{equation}
\mathop{\Delta N_k} \sim \frac{\mathop{\Delta n_\gamma}}{4\pi k_\star^2 /(2\pi)^3}\cdot \mathop{\Delta t}.
\end{equation}
In consequence, the Bose-enhanced regime is reached on a time-scale
\begin{equation}
\frac{1}{\Delta t}\sim \frac{\sqrt{\pi}}{2\sqrt{2}}\,  g_{\phi\gamma\gamma} \Phi\, m_\phi,
\end{equation}
which corresponds to the time-scale for the exponential growth obtained in Eq.~\eqref{eq:growth} using the equations of motion.
The corresponding width $\delta k\sim 1/\Delta t$ parametrically matches Eq.~\eqref{eq:width}, justifying its use above. At later times, when $N_{k}\gtrsim 1$, the width is also of the same parametric size. This is due to the fact that after each subsequent time $\sim\Delta t$, the newly created photons from the stimulated decay dominate the occupation number and ``reinitialize'' the uncertainty principle.

A more careful numerical analysis of the EoM reveals that, for arbitrary initial conditions, the typical time needed to reach the regime of exponential growth is given by this same time-scale.
In this sense, the perturbative calculation and the results from the EoM are consistent.

\section{Expansion  prevents growth}\label{sec:expansion}
As already indicated, so far a crucial effect has been neglected: the expansion of the Universe.
This can be incorporated into the EoM, most conveniently by using conformal time (see Appendix~\ref{app:eom} for the precise expressions). 
However, the effect that is relevant for our purposes is just the fact that the photon $k$-modes are redshifted due to the expansion and therefore move out of the resonance band.
This allows for a physical and intuitive approach to the problem, which we follow in this section.

After a time $\mathop{\Delta t}\ll H$, where $H$ is the Hubble rate of expansion, the physical momentum of the photon modes changes by an amount
\begin{equation}
\frac{\mathop{\Delta k}}{k} \simeq H\mathop{\Delta t} .
\end{equation}
For Bose enhancement to be effective, the exponential growth of the photon number, which is proportional to $|A|^2$, has to set in on a short time-scale compared to the one at which modes are redshifting out of the resonance. This requires
\begin{equation}
\label{eq:nodecay}
{\mathop{\Delta t}}_{\rm growth}\sim \frac{1}{2\eta_{k_\star}} \lesssim \frac{\mathop{\delta k}}{k_\star} \frac{1}{H} \sim {\mathop{\Delta t}}_{\rm redshift}.
\end{equation} 
The actual condition for instability is slightly stronger: the resonance has to be effective for a long enough time to allow the ALP condensate to decay. Decay of the dark matter field happens when a sizeable fraction of its energy density is lost. The energy density in the exponentially growing photon modes is given by
\begin{equation}
\rho_\gamma(t) = \frac{1}{\pi^2} \int \mathop{\diff k}k^2 \omega_k N_k(t),
\end{equation}
where $\omega_k$ is the frequency of a (single helicity) mode and $N_k$ is its occupation number,
\begin{equation}\label{eq:occupation_number}
N_k = \frac{\omega_k}{2}\left( | \dot{A}_k |^2 / \omega_k^2 + \left| A_k \right|^2 \right).
\end{equation}
 Given that the resonance has a very narrow bandwidth $\mathop{\delta k} \ll k$, the integral can be approximated using the method of steepest descent to give
\begin{equation}
\rho_\gamma(t) \simeq \frac{1}{\pi^2}\, k_\star^3 \mathop{\delta k} \sqrt{\frac{\pi}{\eta_{k_\star}t}}\, N_{k_\star}(t),
\end{equation}
where the rate $\eta_{k_\star}$ at centre of the resonance band $k_\star = m_\phi/2$ is given in Eq.~\eqref{eq:growth}, and thus
\begin{equation}
N_{k_\star}(t) = N_{k_\star}^0 \me^{2\eta_{k_\star}t}.
\end{equation}
Here, $N_{k_\star}^0$ denotes the initial occupation number of the mode. As discussed in the previous section, the exponential enhancement sets in when $2\eta_{k_\star}\mathop{\Delta t}\gtrsim 1$. After that point, the ratio of the energy density in the produced photon population with respect to that stored in the ALP field, $\rho_\phi\simeq m^2_\phi \Phi^2 / 2$, can be approximated by
\begin{equation}\label{eq:energy_density_ratio}
\frac{\rho_\gamma}{\rho_\phi} \sim \frac{g_{\phi\gamma\gamma}\,m_\phi^2}{\Phi}\, N_{k_\star}^0 \me^{2\eta_{k_\star}t}.
\end{equation}
The initial condition can correspond to vacuum fluctuations, $N_{k_\star}^{\mathrm{vac}}=1/2$, or to CMB photons with a corresponding occupation 
number
\begin{equation}
N_{k_\star}^{\mathrm{CMB}}=\frac{1}{\exp\left(m_\phi/(2T_{\rm CMB})\right)-1}\sim \frac{2T_{\rm CMB}}{m_\phi},
\end{equation}
valid for $m_\phi\ll T_{\rm CMB}$, using the CMB temperature at the time at which the resonance is active. In both cases, the prefactor of the exponential in Eq.~\eqref{eq:energy_density_ratio} is very small. 
Therefore, for a sizeable fraction of the DM ALPs to decay the resonance has to be active for many growth times,
\begin{equation}
{\mathop{\Delta t}}_{\rm decay} \sim \zeta {\mathop{\Delta t}}_{\rm growth} \sim\zeta/(2\eta_{k_\star}) 
\end{equation}
where
\begin{equation} \label{eq:zeta}
\zeta\sim \mathrm{max}\left\{ \log \left( \frac{\Phi}{g_{\phi\gamma\gamma}m_\phi^2}\right),\, \log \left( \frac{\Phi}{g_{\phi\gamma\gamma}m_\phi T_{\rm CMB}}\right) \right\}.
 \end{equation} 
Thus, the final condition for stability, ${\mathop{\Delta t}}_{\rm decay} \lesssim {\mathop{\Delta t}}_{\rm redshift}$, can be recast as
\begin{equation}
\label{eq:nodecayF}
 g^2_{\phi\gamma\gamma} \Phi^2(t) \lesssim 2\zeta \frac{H(t)}{m_\phi},
\end{equation}
which is required to hold at any time during the evolution of the Universe. $\zeta$ amounts to a factor of $\mathcal{O}(10-100)$ for the parameters of interest.

 As shown in Appendix~\ref{app:eom}, the condition in Eq.~(\eqref{eq:nodecayF}) can also be derived by an  investigation of the EoM for the photon field in a Friedmann--Robertson--Walker (FRW) background. The analysis in Appendix~\ref{app:eom} also shows that other effects related to expansion, namely the damping of the amplitude of the photon field and the ALP oscillations, are not as effective in curbing the exponential growth as the one discussed above. 

The requirement for stability in Eq.~\eqref{eq:nodecayF} must hold at all times.
In Fig.~\ref{fig:master}, the dotted and dashed blue lines correspond to it being evaluated today and at matter-radiation equality, respectively.
We clearly see that the constraint becomes stronger when evaluated at earlier times.
The reason for this is that the left hand side of Eq.~\eqref{eq:nodecayF} falls with the scale factor as $1/a^{3}$, whereas the right hand side falls only as $1/a^2$.
The stability condition is therefore most restrictive when evaluated at the earliest possible time. This corresponds to the time $t_{\mathrm{osc}}$ when the ALP field oscillations start, given by $H (t_{\rm osc}) \sim m_\phi/3$. Assuming standard cosmology, this produces the solid blue line labelled as ``Expansion" in Fig.~\ref{fig:master}.

The QCD axion and simple ALPs that arise as Goldstone bosons inhabit regions which are below the blue line in Fig.~\ref{fig:master}.  These are thus stable even without taking into account plasma effects (this was already understood in the original works~\cite{Abbott:1982af,Preskill:1982cy}). The reason for this is that a sufficient condition for the stability condition in Eq.~\eqref{eq:nodecayF} to be obeyed is for the product $g_{\phi\gamma\gamma} \Phi (t_{\rm osc})$ to be smaller than unity. This is automatically satisfied for simple pseudo-Goldstone bosons, as in this case the field value is bounded by~\cite{Gupta:2015uea}
\begin{equation}
\Phi \lesssim \pi f.
\end{equation} 
At the same time, the photon coupling can be written as
\begin{equation}
g_{\phi\gamma\gamma}=\frac{C_{\phi\gamma\gamma}\alpha_{\rm em}}{4\pi f},
\end{equation}
where $C_{\phi\gamma\gamma}$ summarizes the model dependence (e.g. due to charge assignments of fermions mediating the interaction). In simple models, this constant is not expected to be much larger than unity and therefore $g_{a\gamma\gamma}\Phi\lesssim 1$.
However, there is a renewed interest in ALP models that can accommodate larger values of this product. Recently, it has been realised that this part of the parameter space can be populated by monodromy~\cite{McAllister:2008hb,Jaeckel:2016qjp} or clockwork constructions~\cite{Choi:2014rja,Choi:2015fiu,Kaplan:2015fuy}, and by string theory expectations~\cite{Halverson:2019cmy}. In these situations, the relation $g_{a\gamma\gamma}\Phi\lesssim 1$ is violated by either having an initial field value that is larger than that of a naive Goldstone boson or by directly enhancing the photon coupling over its usually expected value.

\section{Plasma effects prevent early decay}\label{sec:plasma_mass}
There is still a critical observation that we have ignored in our analysis so far: the early Universe is filled with a dense plasma that has a severe impact on the propagation of photons. Following~\cite{Raffelt:1996wa}, the coherent interactions with the background can be described with a modified dispersion relation and a wavefunction renormalization for the photon. 

Importantly, a photon propagating in a medium acquires an extra polarization degree of freedom, that can be identified as a longitudinal excitation in the Lorentz gauge. However, the homogeneous ALP field does not couple to this degree of freedom. This can be seen by explicitly writing down the terms in the photon EoM arising from the $\phi F\tilde{F}$ term in the Lagrangian,
\begin{equation}
\frac{1}{4}g_{a\gamma\gamma} \phi F_{\mu\nu} \tilde{F}^{\mu\nu} \quad \xrightarrow{{\rm EoM}} \quad g_{a\gamma\gamma} \left( \dot{\phi}\, \mathbf{\nabla} \times \mathbf{A} - \mathbf{\nabla}\phi \times (\dot{\mathbf{A}} - \mathbf{\nabla} A_0)  \right),
\end{equation}
where $A = (A^0,\mathbf{A})$ and $A^0$ is completely determined by the Lorentz gauge condition $\partial\cdot A=0$. 
As in this work are only dealing with a spatially homogeneous ALP field, only the first term in the above equation is non-vanishing. This term only sources transverse photons and accordingly we do not further discuss longitudinal modes. 

As discussed in Appendix~\ref{app:plasma_mass}, the most relevant effect\footnote{The wavefunction renormalization constant of the transverse mode is always very close to one and does not play a role in our analysis~\cite{Raffelt:1996wa}.} for our purposes is the fact that the dispersion relation of transverse modes is modified in such a way that the medium does not allow modes with frequency below a certain cutoff to be excited. The lowest frequency that can propagate is known as the \emph{plasma frequency}, $\omega_P$. In practice, this is equivalent to considering an effective in-medium mass\footnote{Strictly, only modes with wavenumber $k\gg \omega_{\mathrm{P}}$ have a dispersion relation that approaches that of a massive particle (see Appendix~\ref{app:plasma_mass} and~\cite{Raffelt:1996wa} for more details). } $m_\gamma = \omega_P$ for the photons. The immediate consequence is that the decay of the ALP cannot take place at all as long as its mass is below the $2m_\gamma$ threshold~\cite{Abbott:1982af}.
This can be seen by adding a term $m_\gamma^2 A_{\pm}$ in the left hand side of the EoM in Eq.~\eqref{eq:motion}, so that the condition for parametric resonance becomes
\begin{equation}
\label{eq:phasespace_expansion}
\left| \sqrt{k^2 + m_\gamma^2} - \frac{m_\phi}{2} \right| < \delta k,
\end{equation}
in terms of physical momenta. If $m_\phi < 2m_{\gamma}$, the effective mass prevents the resonance and thus the decay of the ALP from happening. This effect turns out to be efficient in stabilizing the ALP in a large part of the parameter space.

\begin{figure}[!t]
\centering
\includegraphics[width=0.7\textwidth]{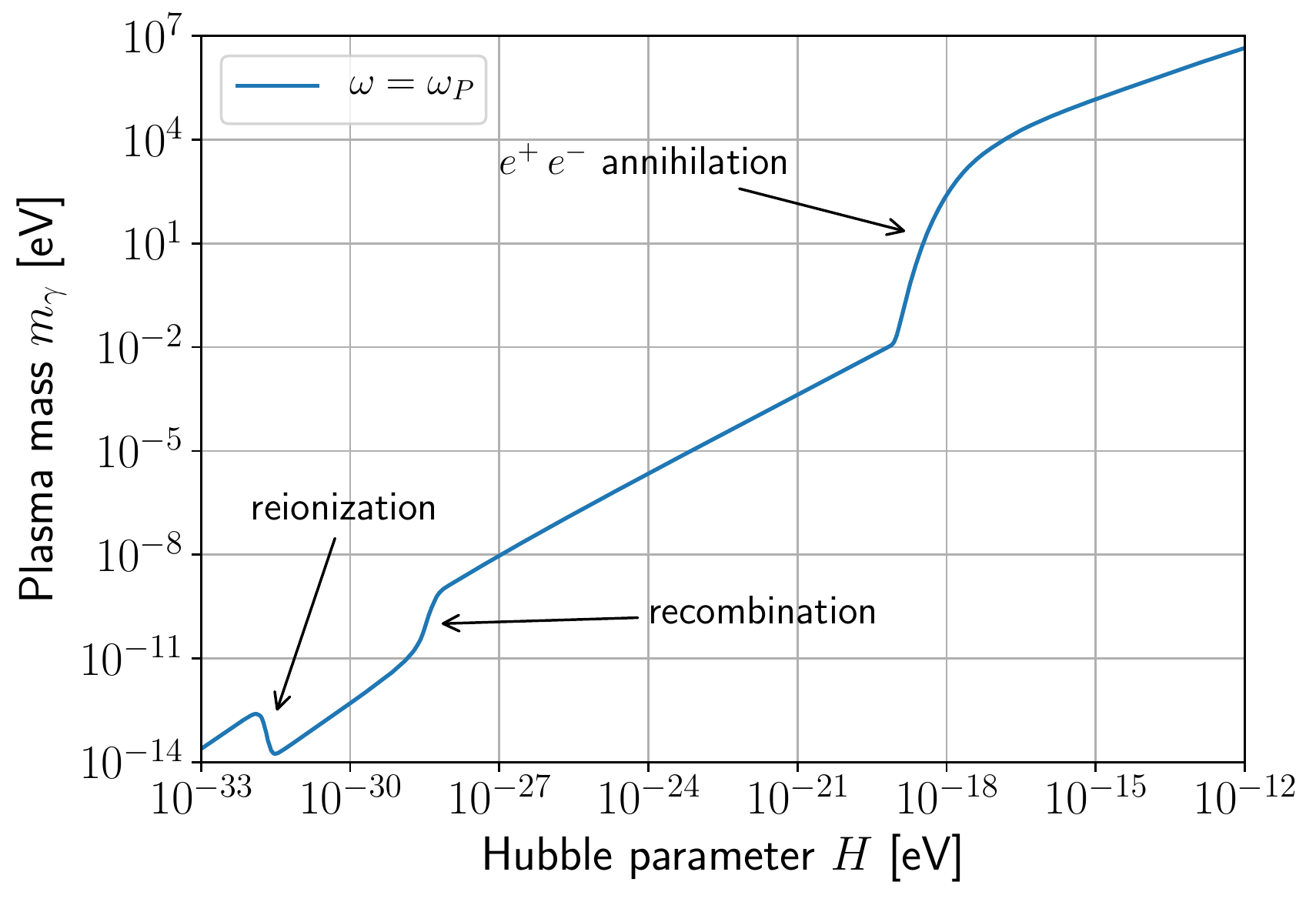}
\caption{Effective mass of the photon (in the sense explained in the text) in the early Universe as a function of the Hubble parameter (see Fig.~\ref{fig:plasma_mass_T_z} for $m_\gamma$ as a function of temperature and redshift). For this figure, we fix $\omega=\omega_P$. 
}
\label{fig:plasma_mass_H}	
\end{figure}

In order to quantitatively evaluate this effect, an expression for the effective photon mass (or \emph{plasma mass}) throughout the relevant epochs of the history of the Universe is required. To obtain it, we use the results of~\cite{Raffelt:1996wa,Braaten:1993jw} and apply them to our setup. The result is presented in Fig.~\ref{fig:plasma_mass_H} as a function of the Hubble parameter, while the corresponding dependencies on the temperature and redshift, together with the details of the calculation, are given in Appendix~\ref{app:plasma_mass}.

The shape of the curves in Figs.~\ref{fig:plasma_mass_H} and~\ref{fig:plasma_mass_T_z}  can be qualitatively understood as follows. At high temperatures $T\gg1\,\mathrm{MeV}$, relativistic electron-positron pairs dominate the dispersion relation of photons and generate a large effective mass $m_\gamma\propto T$. The effective mass sharply drops during the $e^+ e^-$ annihilation at $T\sim m_e$, after which the dominant contribution comes from scatterings off the leftover electrons and the ionized hydrogen and helium atoms. At even lower temperatures $T\sim 0.1\,\mathrm{eV}$, electrons and protons combine to form neutral atoms causing another drop in the effective mass. Finally, the first stars reionize the medium at around $z\sim 1-10$, which explains the increase in the plasma mass at $T\sim \mathrm{meV}$. Today, the effective photon mass in the intergalactic medium\footnote{The space between galaxies is thought to be permeated by an ionized plasma with electron densities $n_e\sim 10^{-6}-10^{-4}\,\mathrm{cm}^{-3}$ and temperatures $T\sim 10^{5}-10^{7}\,\mathrm{K}$, known as the warm-hot intergalactic medium or WHIM~\cite{Nicastro:2018eam}.} is expected to be $m_\gamma\sim 10^{-14}-10^{-13}\,\mathrm{eV}$, which sets an absolute lower bound on the mass of the (background) ALPs that can be subject to decay.

As soon as the plasma is dilute enough such that $m_\gamma < m_\phi / 2$, the ALP decay is allowed and the discussion in the previous section becomes valid. Given that the condition in Eq.~\eqref{eq:nodecayF} is stronger earlier on, it must be imposed at the first moment that it becomes applicable. Two requirements are needed: that the ALP field is oscillating (i.e. that $3H \lesssim m_\phi$) and that the photon plasma mass falls beyond the threshold $m_\gamma \lesssim m_\phi / 2$. By looking at Fig.~\ref{fig:plasma_mass_H}, one immediately concludes that the latter is the most restrictive of the two. This means that it is the plasma effects that set the earliest time at which the resonance can be active, and ultimately the region of parameter space where the ALP dark matter field is cosmologically unstable.

Numerically, the region of ALP parameter space where Bose enhancement can be relevant is delineated by the red line labelled ``Plasma" in Fig.~\ref{fig:master}. It is obtained by imposing Eq.~\eqref{eq:nodecayF} as soon as $2 m_\gamma < m_\phi$. As discussed above, there is no bound on  ALPs lighter than present-day photons in the intergalactic medium, i.e. for $m_\phi \lesssim 10^{-14}\,\mathrm{eV}$. 
We conclude that when plasma effects are considered, the region that is prone to instabilities is much smaller due to the delay of the ALP decay. As advertised in the introduction, it is therefore crucial to take into account both the expansion of the Universe and the in-medium dispersion effects to absolutely establish the stability of ALP dark matter.

\bigskip

Let us conclude this section by commenting on the possibility of photon production due to spinodal or tachyonic instabilities~\cite{Agrawal:2017eqm, Lee:1999ae}. As soon as the temperature is low enough such that the photon plasma mass can be ignored, one can see from Eq.~\eqref{eq:motion} and Eq.~\eqref{eom2} that the frequency of certain modes of the photon field can become imaginary if they lie in the band $\lambda- 2q <0$.
The mode remains tachyonic only as long as the sign of the cosine does not change. Requiring that the timescale of the energy growth is smaller than the time that the cosine preserves its sign, in addition to the tachyonic condition, gives~\cite{Agrawal:2017eqm}
\begin{equation}
m_\phi< \frac{k}{a}<g_{\phi\gamma \gamma}\Phi m_\phi,
\end{equation}
 which clearly requires $g_{\phi\gamma \gamma}\Phi >1$. It is easy to see that the condition for photon production due to a tachyonic instability is not satisfied anywhere in the stable region below the red line in Fig.~\ref{fig:master}. Indeed, from Eq.~\eqref{eq:nodecayF}, we see that  the product $g_{\phi\gamma \gamma}\Phi$ is at most of the order of $\sqrt{H/m_\phi}$, which is much smaller than unity by the time the photon mass becomes smaller than the ALP mass (see Fig.~\ref{fig:plasma_mass_H}).

\section{Conclusions}\label{sec:conclusions}
Axion-like particles (ALPs) are dark matter candidates that often enjoy a $\phi F\tilde{F}$ coupling, through which the ALP can decay into two photons. That said, the cosmological stability of this kind of dark matter seems to be nearly trivial at first sight due to its low mass and incredibly weak coupling. However, if the ALP field exists already during inflation, the rapid expansion creates an extremely homogeneous and therefore coherent initial state. Given the low mass of the ALPs, this effectively corresponds to a bosonic state with incredibly high occupation numbers, suggesting the possibility of a Bose enhancement of the dark matter decay. Indeed, in an otherwise empty and non-expanding Universe, the resulting stimulated emission would lead to an extremely rapid decay of ALPs into photons, invalidating most of the interesting parameter space. However, as already found in the original papers~\cite{Abbott:1982af,Preskill:1982cy,Dine:1982ah}, there are two effects that prevent this catastrophe: the expansion of the Universe and the plasma effects that modify the propagation of photons. For QCD axions, the former is already sufficient to prevent the evaporation of the dark matter; however, for ALPs with larger couplings significant non-excluded areas of parameter space would still be subject to decay. Here the plasma effects are crucial, as they generate an effective photon mass which kinematically forbids the decay of ALPs lighter than this plasma mass. This opens up the full parameter space not yet excluded by past experiments or astrophysical observations (which do not rely on ALPs being the dark matter). In this sense, the full region is viable to be explored by current and near future ALP dark matter experiments~\cite{Kahn:2016aff,Shokair:2014rna,brasscite,Petrakou:2017epq,Droster:2019fur,Gatti:2018ojx,Majorovits:2017ppy,McAllister:2017lkb,Alesini:2019ajt,Melcon:2018dba}.  

It may be interesting to think of additional observable effects caused by the Bose enhanced decay stimulated by the CMB photons. Even if the ALP condensate is not significantly depleted, this leads to an additional number of photons of frequency $\sim m_{\phi}/2$ being injected. Usually this effect is bigger at relatively small masses where, however, it seems challenging to detect. The interplay of the condensate with primordial magnetic fields can also lead to interesting phenomena~\cite{Ahonen:1995ky}. Although first estimates suggest that the cosmological stability of the ALPs is not threatened under standard assumptions regarding the magnetic field strength, this is a topic worth studying in more detail.

For ALP couplings other than the photon one, similar Bose enhanced effects may become important and therefore the viability of the ALP as a DM candidate needs careful verification. Indeed, while self-couplings have been found to be safe for QCD axions~\cite{Abbott:1982af,Kolb:1998iw}, they can be relevant in models that go beyond the simplest pseudo-Goldstone ALP. Among other effects, they can lead to a fragmentation of the condensate~\cite{Berges:2019dgr}\footnote{The fragmentation does not invalidate these light scalars as dark matter candidates, but can significantly modify their cosmological and astrophysical phenomenology.}. Similarly, some care is needed when couplings, e.g. to other light dark sector bosons, are stronger than expected for naive pseudo-Goldstone bosons as suggested by recent theoretical constructions~\cite{McAllister:2008hb,Jaeckel:2016qjp,Choi:2014rja,Choi:2015fiu,Kaplan:2015fuy,Halverson:2019cmy}.

\section*{Acknowledgments}
We thank S.~Davidson for very useful discussions and J.~Redondo for sharing his wisdom and his personal notes on the topic with us.
JJ would like to thank the IPPP for hospitality during the time this project was started and support within the DIVA fellowship programme.
GA is a grateful recipient of a ``la Caixa" postgraduate fellowship from the Fundaci\'on ``la Caixa".

\appendix

\section{Equation of motion of the photon in an expanding Universe}
\label{app:eom}
The condition in Eq.~\eqref{eq:nodecayF} can be derived by investigating the photon EoM in an FRW background, 
\beq
\frac{\mathop{\diff^2 A}}{\mathop{\diff x^2}} + \frac{2 H}{m_\phi} \frac{\mathop{\diff A}}{\mathop{\diff x}}+\(\lambda \frac{1}{a^2} - 2 q \frac{1}{a^{5/2}}  \cos(2 x)\) A=0,
\label{eom2}
\eeq
where we set the scale factor $a$ to be $1$ at $t=t_0$, and define $x=m_{\phi}t/2$, $\lambda= 4 k^2 / m_\phi^2$ and $q= 2 g_{\phi\gamma \gamma} \Phi k / m_\phi$ such that they coincide with their flat space definitions. Taking  $a\sim t^n\sim x^n$, with  $n=1/2$ (2/3) for radiation (matter) domination, we can replace factors of $1/a$ by factors of $(x_0/x)^n$, where $x_0=x(t_0)$. 

 Here, we focus on the regime where the photon plasma mass is much smaller than the ALP mass, i.e. $m_\gamma \ll m_\phi$, so that the exponentially enhanced decay is possible. From Fig.~\ref{fig:plasma_mass_H}, we see that this implies $x=m_\phi t /2=n m_\phi/(2 H) \gg1$.   We thus choose an initial time point such that $x_0\equiv1/\xi\ll1$, and make a change of variables $y= x-x_0$  to obtain
\beq
\frac{\mathop{\diff^2 A}}{\mathop{\diff y^2}} + \(\frac{\xi n}{1+\xi y}\)^{2 n} \frac{\mathop{\diff A}}{\mathop{\diff y}} +\(\lambda\(\frac{1}{1+\xi y}\)^{2 n}- 2 q \(\frac{1}{1+\xi y}\)^{{5 n}/2} \cos(2 y+ \delta)\) A=0.
\label{xeq}
\eeq
 Note that for $\xi=0$ we recover the Mathieu equation in Eq.~\eqref{mathieu}. For non-zero $\xi$,  exponential growth may not take place if, as explained above, the $k$-modes get redshifted out of the unstable band.  The redshifting causes  $\lambda /a^2$ to decreases with time as
\beq
\lambda\(\frac{1}{1+\xi y}\)^{2 n}\approx  \lambda \(1- 2n\xi y\),
\eeq
so that the time it takes for $k$-modes to move out of the unstable band  $|\lambda - 1| < q$ is
\beq
\Delta y_{\rm redshift}= \frac{q}{2n \xi}.
\eeq
Demanding this to be smaller than the time scale of exponential growth, $ \Delta y_{\rm decay}=  \zeta/q$, we again find
\beq
q^2 \lesssim  2 \zeta n  \xi \quad \Rightarrow \quad (g_{\phi\gamma \gamma} \Phi)^2 \lesssim 2 \zeta \frac{H}{ m_\phi}.
\eeq
where $\zeta$ is defined in  Eq.~\eqref{eq:zeta}.

It is worth noting that there are  two other effects that can potentially curb the exponential enhancement. The first is the Hubble friction (second term in Eq.~\eqref{eom2}) and the second is the damping of the ALP oscillations (that is, the decrease in $q / a^{5/2}$ with time). These effects, however, are subdominant compared to the redshifting discussed above. While the first effect becomes relevant only for $\xi \sim q$, the second is not important because the change in $q$ in the time $\Delta t_{\rm redshift}$ is ${\cal O}(q^2)$ and thus marginal.

\section{Dispersion relation of the photon in the primordial plasma}\label{app:plasma_mass}
Here we briefly review the most relevant dispersion effects that affect the propagation of photons in the early Universe plasma. The discussion is based on~\cite{Raffelt:1996wa,Braaten:1993jw} whom we closely follow. Other references where similar studies are made are~\cite{Redondo:2008ec, Mirizzi:2009iz, Dvorkin:2019zdi}, among others.

A photon propagating through a plasma is subject to scattering with the charged constituents of the medium. The coherent accumulation of such interactions can be macroscopically described by a modification of the dispersion relation of the photon, which becomes
\begin{equation}\label{eq:dispersion_relations}
-\omega^2 + k^2 + \pi_a(\omega, k) = 0.
\end{equation}
The index $a=\{T(\pm), L\}$ labels the different polarizations of the field, now also including a longitudinal degree of freedom that is not present in vacuum. Thus, the key quantities are $\pi_L$ and $\pi_T$, the longitudinal and transverse components of the polarization tensor $\Pi^{\mu\nu}$.
In an ionized plasma, the dominant contribution to the polarization tensor is that of the free electrons and positrons. 
At lowest order in QED~\cite{Raffelt:1996wa,Braaten:1993jw},
\begin{eqnarray}
\pi_L &=& \omega_P^2 \left( 1 - G(v_\star^2k^2/\omega^2) \right) + (v_\star^2-1) k^2, \\
\pi_T &=& \omega_P^2 \left( 1 + G(v_\star^2k^2/\omega^2)/2 \right),
\end{eqnarray}
with 
\begin{equation}
G(x) = \frac{3}{x} \left( 1 - \frac{2x}{3} - \frac{1-x}{2\sqrt{x}}\log\left( \frac{1+\sqrt{x}}{1-\sqrt{x}} \right) \right).
\end{equation}
In the above expression, $v_\star = \omega_1 / \omega_P$ depends on the \textit{plasma frequency} $\omega_{P}$ and  $\omega_1$, 
\begin{eqnarray}\label{eq:plasma_frequency}
\omega_P^2 &=& \frac{4\alpha_{\rm em}}{\pi} \int_0^\infty \mathop{\diff p} f_p\, p\, \left(v - \frac{1}{3} v^3 \right) , \\
\omega_1^2 &=& \frac{4\alpha_{\rm em}}{\pi} \int_0^\infty \mathop{\diff p} f_p\, p\, \left(\frac{5}{3} v^3 - v^5 \right) ,
\end{eqnarray}
where $f_p$ is the sum of the phase-space distributions of $e^{-}$ and $e^{+}$. 

For nondegenerate systems, as appropriate in the early Universe, one has
\begin{equation}
\omega_P^2 \simeq \begin{cases} 
\frac{4\pi\alpha_{\rm em} n_e}{m_e} \left( 1- \frac{5}{2} \frac{T}{m_e} \right) &\qquad\mathrm{Nonrelativistic},\\ 
\frac{4\alpha_{\rm em}}{3\pi} \left( \mu^2 + \frac{1}{3}\pi^2T^2 \right) &\qquad\mathrm{Relativistic}, \end{cases} 
\end{equation}
where $T$, $n_e$ and $\mu$ are the temperature, number density and chemical potential of the electrons, respectively. This can be inserted into the dispersion relation, Eq.~\eqref{eq:dispersion_relations}.

In practice, we are only interested in the cutoff frequency below which no propagating modes (i.e., modes with non-vanishing real $k$) exist in the plasma. We focus on transverse modes, as these are the only ones that couple to the homogeneous ALP field\footnote{Longitudinal modes have a more complex behaviour that may be interesting to study in inhomogeneous situations.}.
Given that $\pi_T$ is a monotonically increasing function of $k$, we can simply solve Eq.~\eqref{eq:dispersion_relations} for $k=0$, in which case $G=0$. It is thus easy to see that the cutoff frequency is straightforwardly given by the plasma frequency $\omega_{P}$.
In analogy to the case of a massive particle, for which the mass gives a lower bound for the frequency (or energy) of the propagating states, we refer to the cutoff frequency as the \emph{plasma mass} of the photon, and denote it by $m_\gamma$. Note, however, that this is an abuse of nomenclature, as the dispersion relation for transverse modes only really approaches that of a massive particle for short wavelengths $k\gg\omega_P$. Our definition of $m_\gamma$ differs slightly from the standard one (e.g.~\cite{Raffelt:1996wa}), by up to a factor of $3/2$ in the relativistic limit.

\begin{figure}[!t]
\centering
\includegraphics[width=0.95\textwidth]{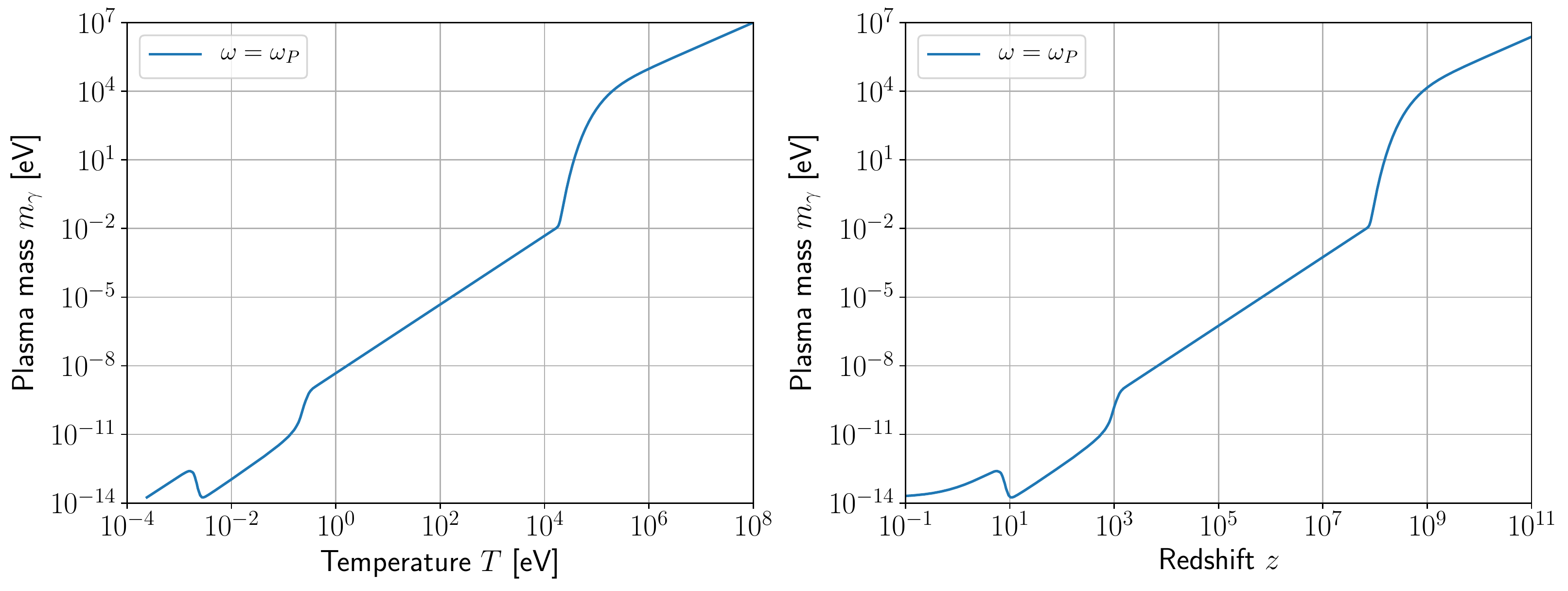}
\caption{Effective transverse photon mass in the early Universe as a function temperature (\textit{left}) and redshift (\textit{right}) (see Fig.~\ref{fig:plasma_mass_H} for $m_\gamma$ as a function of the Hubble parameter). For this figure, we fix the frequency to the plasma one, $\omega = \omega_P$.}
\label{fig:plasma_mass_T_z}	
\end{figure}

    At very early times in the history of the Universe, the plasma is comprised primarily of $e^{+}e^{-}$ pairs (before $e^{+}e^{-}$ annihilation) and ionized hydrogen atoms (after $e^{+}e^{-}$ annihilation). This entails that the interactions of photons with the medium are indeed dominated by Coulomb scattering with electrons and positrons. However, at lower temperatures the protons and electrons recombine to form neutral atoms and the free electron density becomes very small. As a result, the scattering off hydrogen and helium atoms cannot generally be neglected anymore and there is an additional contribution to the transverse photon mass, which becomes~\cite{Mirizzi:2009iz}
\begin{equation}
m_\gamma^2 \simeq \omega_P^2 - 2\omega^2 (\mathrm{n}-1)_{\mathrm{H}} - 2\omega^2 (\mathrm{n}-1)_{\mathrm{He}}.
\end{equation}
Here, $\mathrm{n}_X\equiv\mathrm{n}_X(n_X)$ is the refractive index, which is proportional the density of atoms of the species $X$. The contribution from helium can be neglected because it is less abundant (it has a mass fraction $Y_p=m_{\rm He}/m_{\rm H}\simeq 0.25$) and its refractive index, $(\mathrm{n}-1)_{\mathrm{He}}=0.36\times 10^{-4}$, is much smaller than the hydrogen one, $(\mathrm{n}-1)_{\mathrm{H}}=1.32\times 10^{-4}$ (values for normal conditions~\cite{Optics5thEd}). The transverse photon mass can then be given as a function of the proton density $n_p$ and the ionization fraction $X_e$ as
\begin{equation}\label{eq:plasma_mass_atoms}
m_\gamma^2 = 1.4\times 10^{-21}\,\mathrm{eV}^2\,\left( X_e - 7.7\times 10^{-3}\left( \frac{\omega}{\mathrm{eV}} \right)^2 (1-X_e)  \right) \frac{n_p}{\mathrm{cm}^{-3}}.
\end{equation}
One may worry that during the dark ages, the negative contribution from the scattering off neutral hydrogen can drive $m_\gamma^2$ to be negative for high-frequency photons. This does however not play any role in our analysis, given that the photons produced from resonant ALP decays have frequencies only slightly above $\omega_P$, too small for the second term in Eq.~\eqref{eq:plasma_mass_atoms} to be relevant.
Putting everything together, we obtain the profile for the cosmological history of the plasma mass of the photon shown in Fig.~\ref{fig:plasma_mass_T_z}, both as a function of temperature and redshift (Fig.~\ref{fig:plasma_mass_H} shows the corresponding line as a function of the Hubble parameter). For the calculation, we fix $\omega=\omega_P$ and use the ionization fraction profile from~\cite{Mirizzi:2009iz}.

\bibliographystyle{utphys}
\bibliography{references}

\end{document}